\documentclass[prb,twocolumn,superscriptaddress,altaffillsymbol]{revtex4-1}
\usepackage{amssymb}
\usepackage{amsmath}
\usepackage{commath}
\usepackage{graphicx}
\usepackage{verbatim}
\usepackage{epsfig}
\usepackage{epstopdf}
\usepackage{color}
\usepackage{siunitx}
\usepackage{array}
\usepackage{upgreek}
\newcolumntype{P}[1]{>{\centering\arraybackslash}p{#1}}
\newcolumntype{M}[1]{>{\centering\arraybackslash}m{#1}}

\DeclareSIUnit\gauss{G}

\begin{document}

\title{Anomalously small superconducting gap in the strong spin-orbit coupled superconductor: $ \beta- $Tungsten}

\author{Prashant Chauhan}
\affiliation{Department of Physics and Astronomy, The Johns Hopkins University,  3701 San Martin Dr., Baltimore, Maryland 21218, USA}

\author{Ramesh Budhani}
\affiliation{ Department of Physics, Morgan State University, 1700E Cold Spring Lane, Baltimore, Maryland 21251, USA}

\author{N.~P.~Armitage} 
\affiliation{Department of Physics and Astronomy, The Johns Hopkins University,  3701 San Martin Dr., Baltimore, Maryland 21218, USA}

\begin{abstract}
Thin films of $ \beta- $tungsten host superconductivity in the presence of strong spin-orbit coupling.  This non-equilibrium crystalline phase of tungsten has attracted considerable attention in recent years due to its giant spin Hall effect and the potential promise of exotic superconductivity. However, more than 60 years after its discovery, superconductivity in this material is still not well understood. Using time-domain THz spectroscopy, we measure the frequency response of the complex optical conductivity of $ \beta- $tungsten thin film with a T$_c$ of 3.7 K in its superconducting state. At temperatures down to 1.6 K, we find that both the superconducting gap and the superfluid spectral weight are much smaller than that expected for a weakly coupled superconductor given the T$_c$.  The conclusion of a small gap holds up even when accounting for possible inhomogeneities in the system, which could come from other crystalline forms of tungsten (that are not superconducting at these temperatures) or surface states on $ \beta- $tungsten grains. Using detailed X-ray diffraction measurements, we preclude the possibility of significant amount of other tungsten allotropes, strongly suggesting the topological surface states of $ \beta- $tungsten play the role of inhomogeneity in these films. Our observations pose a challenge and opportunity for a theory of strongly anisotropic normal metals with strong spin-orbit coupling to describe.
\end{abstract} 

\maketitle
\setlength\belowcaptionskip{-3ex}

Bulk crystalline tungsten (W) in the most stable $ \alpha- $W bcc form has a very low superconducting T$_c $ of 11 mK~\cite{Gibson_PRL_1964}.  However, thin films of W can have  T$_c $'s as large as 5 K e.g. two orders of magnitude higher than the bulk. This has been attributed to the presence of a metastable A15 $ \beta$ phase structure that can be stabilized in thin films~\cite{Basavaiah_JAP_1968, HOFER_TSF_2019}.  Such $\beta-$W has very distinctive mechanical, electrical and optical properties. Its room temperature resistivity ($ \sim$$\SI{200}{\mu\Omega\centi\meter}$) is much higher than that of $ \alpha- $W ($\sim$$\SI{20}{\mu\Omega\centi\meter}  $)~\cite{Hao_APL_2015}.  Due to strong spin orbit coupling (SOC), $ \beta- $W exhibits a giant spin Hall effect with large spin Hall angle ($ \theta_{SH}\sim -0.45 $), making it potentially useful in spintronic applications\cite{Mondal_PRB_2017,Demasius_NatCom_2016}. DFT calculations show that $ \beta- $W may be a Dirac system that hosts massive Dirac fermions and may host surface arc states with novel spin textures reminiscent of those in topological insulators~\cite{LI_PRB_2019,Xie_IEEE_2018}. The helical spin-polarized electrons in topological insulators can host exotic excitations like Majorana fermions if their topological surface states become superconducting by proximity effect~\cite{Mellnik_nat_2014}.  These facts prompt us to investigate the possibility of exotic bulk and surface superconductivity in $ \beta- $W.

The study of superconductivity in $ \beta- $W thin films has proven to be challenging.  There have been only a few experimental and theoretical investigations~\cite{Gibson_PRL_1964, Bond_PRL_1965, Basavaiah_JAP_1968} that have given understanding of its superconducting state and order parameter. This is mainly because of the difficulty in growing clean $\beta-$W thin films without impurity phases like $ \alpha-$W and the instability of this metastable phase at room temperature where it can spontaneously transform into the $\alpha-$W phase~\cite{Xiao_nano_2013}.  However, there has been a long-term (and now renewed) interest in both the metallic state and the superconductivity of $ \beta- $W.  The superconductivity exhibits a number of unusual aspects.  Using tunneling, Basavaiah \textit{et al} showed that the temperature dependence of the superconducting energy gap $ \Delta(T) $ follows the Bardeen-Cooper-Schrieffer (BCS) dependence~\cite{Basavaiah_JAP_1968}, but with a reduced ratio $ \alpha = \Delta/k_{\mathrm{B}}$T$_c \sim 1.1-1.8$, (where $\Delta$ is the superconducting energy gap) as opposed to the universal weak coupling BCS value of $ \alpha_{\mathrm{BCS}} \sim 1.76$~\cite{Basavaiah_JAP_1968,BCS_1957}. It was suggested that the reduced gap could be related to the presence of $ \alpha-$W phase~\cite{Basavaiah_JAP_1968}.   We shall revisit this possibillity below.

In general, the $ \alpha = \Delta/k_{\mathrm{B}}$T$_c$ ratio provides important insight into the physics.  In a clean system, the coupling strength of superconductivity can be defined relative to $\alpha_{\mathrm{BCS}}$, with $ \alpha\simeq\alpha_{\mathrm{BCS}} $ for a weakly coupled superconductor and $ \alpha $ somewhat greater than $\alpha_{\mathrm{BCS}} $ for a strongly coupled superconductor.  In disordered systems, pair breaking tends to decrease T$_c$ faster than it does the $\Delta$, which increases the ratio above $ \alpha_{\mathrm{BCS}} $~\cite{Khasanov_PRB_2021}. Thus, in general we expect $ \alpha \ge\alpha_{\mathrm{BCS}} $~\cite{Khasanov_PRB_2021}, which is in contrast to observations in $ \beta- $W~\cite{Basavaiah_JAP_1968}. However, in such considerations, there is an implicit assumption that the order parameter is uniform both in space and in momentum.  The order parameter can be anisotropic in momentum space and the gap on the Fermi surface $\langle\alpha\rangle$ can be at points smaller than $ \alpha_{\mathrm{BCS}} $~\cite{Khasanov_PRB_2021}. Due to a complicated and non-uniform Fermi surface~\cite{LI_PRB_2019}, we expect $ \beta- $W to possesses  an anisotropic superconducting gap~\cite{carbotte1977microscopic}. However, the question is to what degree?  Real space inhomogeneity can of course create regions where the gap is suppressed and this may also be an effect.

\begin{figure}
	\includegraphics[width=7.5 cm]{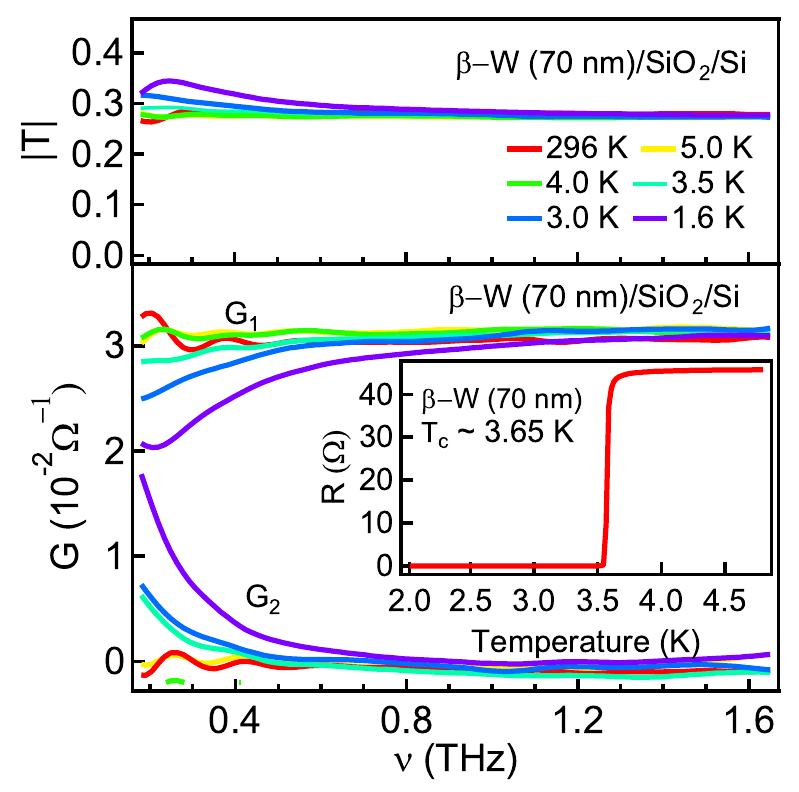}
	\caption{\label{fig:zerofield}(Top) Zero-field magnitude of transmission, (Bottom) Real $G_1 (\nu)$ and imaginary $G_2(\nu)  $  parts of the zero-field complex conductance of $\beta$-W 70 nm thin film  grown on Si substrate as a function of frequency from room temperature (T$\gg$T$_c $) to ($\SI{1.6}{\kelvin}\ll$T$_c$). Inset: Temperature dependence of the 4-probe dc resistance.}
\end{figure}

\begin{figure}
	\includegraphics[width=7 cm]{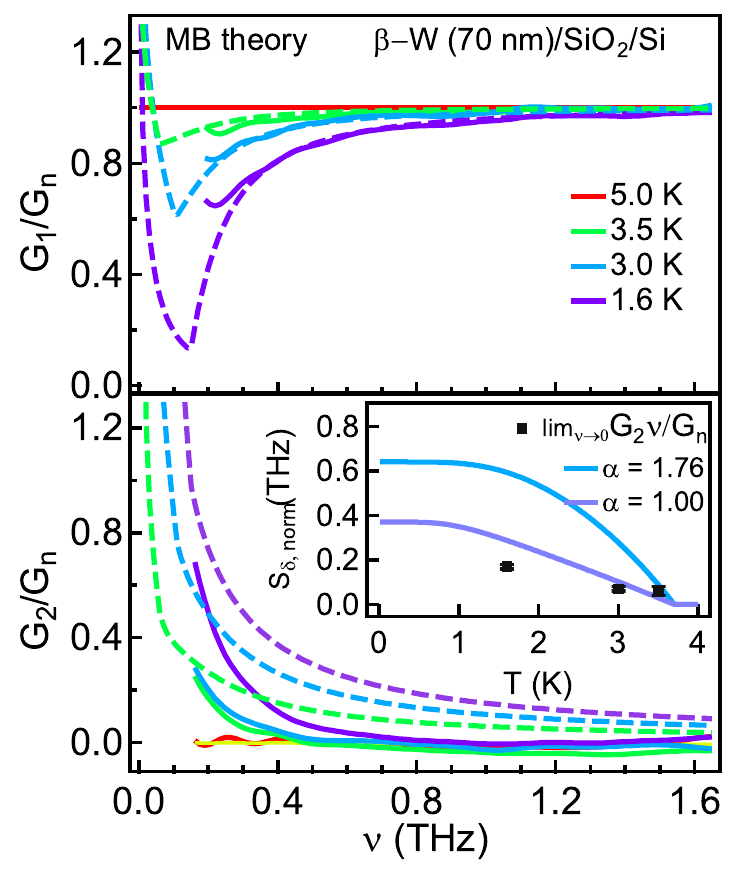}
	\caption{\label{fig2} Real $G_1 (\nu)$ and imaginary $G_2(\nu)  $  parts of the zero-field complex conductance of $\beta$-W 70 nm thin film, normalized to the normal-state conductance $ G_n $. The solid lines are experimental data normalized w.r.t the conductance at 5 K. The dashed lines are fit to the data using Mattis-Bardeen theory for BCS superconductor. Inset:  Black squares show the temperature dependence of the superfluid spectral weight $ S_\delta(T) $ determined from lim$ _{\nu\rightarrow0}\nu G_2/G_n $. The blue line shows the expected spectral weight for a weakly coupled BCS superconductor ($ \alpha_{BCS}=1.76 $). The purple line shows the expected spectral weight for a BCS-like superconductor but with the experimentally determined $ \Delta(0)=0.32$ meV that gives $ \alpha=1 $.} 
\end{figure}

Here we use high precision time-domain THz spectroscopy (TDTS) to measure the low-energy complex optical conductivity of $ \beta- $tungsten films as a function of temperature.   We systematically study and track the superconducting gap, $ \Delta(T) $, as a function of temperature. We find that the superconducting energy gap and the superfluid density can be described phenomenologically in terms of BCS theory, however, with a small energy gap parameter.  This confirms a value of $ \alpha $ much smaller than the 1.76 than expected for a weakly coupled BCS superconductor.   The conclusion of a small gap holds up even when using extant effective medium models that account for inhomogeneity in the form of normal metal inclusions.  Such inclusions could come from other crystalline forms of tungsten (not superconducting at these temperatures) or surface states on $ \beta- $Tungsten grains.  Our observations pose a challenge and opportunity for a theory of strongly anisotropic normal metals with strong spin-orbit coupling to describe.

Thin films of the $ \beta- $Tungsten (A15 structure) were grown on 0.5 mm thin Si(001) with $ \sim200 $ nm thick thermal oxide and 0.5 mm MgO(100) substrates using sputtering in presence of N$ _2 $:Ar gas mixture at room temperature. The Ar sputtering pressure was kept at $ 5 $ mTorr during the deposition. The fact that $\beta-$W thin films require N$ _2 $:Ar or O$ _2 $:Ar mixture  assisted growth has made it difficult to tune the quality and T$ _c $ of the superconducting film. TDTS measurements were performed in transmission geometry on thin films of varying thicknesses ranging from 70 - 130 nm. The 70 nm $ \beta- $W thin-films had the highest T$_c $'s of $ \sim 3.7 $ K. The samples on both Si and MgO substrates gave similar results (See Supplementary Material\cite{Supplementary,Garnett_1904,Tanner_PRB_2013,Carr_academic_1985,Granqvist_PRB_1977,Sihvola_Hindawi_JN_2007} (SM)). Both real and imaginary parts of the complex conductance, $ \tilde{G}(\nu) $, were extracted from the c omplex transmission measured in TDTS measurements, performed down to 1.6 K~\cite{Prashant_PRL_2019}.

Fig. ~\ref{fig:zerofield} shows the frequency dependent transmission and complex conductivity $ \tilde{G}(\nu) $ of a 70 nm $ \beta- $W film, between $ 0.2 - 2  $ THz in zero magnetic field at a few different temperatures above and below T$_c $.   For  T$>$T$_c $, the real part of the conductance, $ G_1(\nu) $ is constant in both this frequency and temperature range with RRR $ \sim1 $.  In the normal state, $ G_2(\nu) $ is zero in correspondence with the large scattering rate ($ \gg$10 THz). The THz conductance in the normal state $ 0.03$ $\Omega^{-1}$ matches well with DC conductance measurement, $ G_{dc} = 0.028$ $\Omega^{-1}$. Below T$_c $ both $ G_1(\nu) $ and $ G_2(\nu) $ shows features indicative of the opening of a superconducting energy gap. As the temperature falls below T$_c $, a depletion develops in $ G_1(\nu) $ at low $ \nu $, corresponding to shift of superconducting carrier spectral weight to the zero frequency delta function~\cite{Tinkham_PR_1968}.  However, the depletion is not large and even at the lowest temperatures of 1.6 K, the conductance remains $ \sim66\% $ of the normal state at $ 0.2 $ THz.  For a weak coupling BCS superconductor with T$_c \sim 3.7 $ K, one expects an optical $2\Delta $ gap of about 0.27 THz at this temperature.  $ G_1(\nu) $ shows a small upturn below 0.2 THz.  As $ \nu \rightarrow0$, $ G_2(\nu) $ shows $ 1/\nu $-like dependence at the lowest temperatures, characteristic of the superconducting state.

To understand the superconducting state better and determine the energy gap $ \Delta(T) $, we simultaneously fit the normalized $ G_1(\nu) $ for {\it all} temperatures to the Mattis-Bardeen (MB) theory~\cite{Mattisbardeen_PR_1958, Prashant_PRL_2019, Tinkham_2nd}. Normalizing the low temperature conductance with the normal state conductance, $ G_1(5 \text{K}) $ eliminates a number of the systematic errors in the transmission data and reduces the number of fitting parameters. For the fitting procedure, the only free parameter is the zero temperature superconducting gap, $ \Delta(0) $. The result of the fits as well as the normalized conductance data are shown as dashed and solid lines in Fig. ~\ref{fig2}.  The upturn in $ G_1(\nu) $ at 0.2 THz becomes more apparent after normalizing the data. The global fit to the real part of the normalized conductance for all temperatures gives $ \Delta(0) = 0.32 $ meV, which is similar to values obtained from tunneling spectroscopy (0.31 - 0.52 meV for films with T$_c  $ ranging from 3.1 K to 3.3 K)~\cite{Basavaiah_JAP_1968}. There is close agreement between the MB fits and $ G_1(\nu) $.   In contrast, the correspondence with the imaginary part using the same parameters as the real part gives poor agreement at low frequencies [Fig. ~\ref{fig2}].  Nevertheless, the temperature evolution of the superconducting gap $ \Delta(T) $, follows standard BCS $ \Delta(T) = \Delta(0)$tanh$[1.74\sqrt{\mathrm{T}_c/\mathrm{T} -1}] $, given for a weakly coupled BCS superconductor. However the extracted gap $ 2\Delta(0) = 0.15$ THz (0.64 meV) or $ \alpha=\Delta(0)/k_{\mathrm{B}}$T$_c \sim 1$ is much less than the weak coupling limit of 1.76 for a fully gapped BCS superconductor.

To get further insight into the superconducting gap and confirm the Mattis-Bardeen fits, we study the temperature dependence of the superfluid spectral weight, $ S_\delta(T) $, as a measure of superfluid density $ n_s $.  We calculate $S_{\delta,G_2} = \text{lim}_{\nu\rightarrow0} \nu G_2/G_n  $ which is a measure of superfluid spectral weight determined directly from the TDTS experimental data. We can compare it to the value calculated $ S_{\delta,\text{MB}} $ for $ \alpha = 1 $ and $ \alpha_{\text{BCS}}=1.76 $ using the approximation for a fully gapped BCS superconductor given by $S_\delta(T)=\textstyle\frac{S_\delta(0)\Delta(T)}{\Delta(0)}\tanh[\Delta(T)/2k_{\mathrm{B}}T] $~\cite{Tinkham_2nd}. The normalization constant $ S_{\delta}(0) $ is given by the Ferrell-Glover-Tinkham (FGT) sum rule, $ S_\delta(0) = S_n -S_{qp}(0)$, where $ S_n $ is the total spectral weight in normal state and $ S_{qp}(0) $ is the above gap spectral weight at $ T\sim0 $ as given by the MB theory~\cite{Prashant_PRL_2019}.  In Fig. ~\ref{fig2}(inset) we compare the temperature evolution of $S_{\delta,G_2}$ with both $ S_{\delta,\text{MB}} $ for $ \alpha=1 $ and $ \alpha_{\text{BCS}}=1.76 $. Near T$_c $, the curve for  $ \alpha=1 $ is close to the experimental value, but over estimates the spectral weight found in the delta function.    Thus we conclude that the Mattis-Bardeen fits below 0.2 THz do not match the actual conductance $ G_1(\nu) $ and $ S_{\delta,\text{MB}}(T) $ at low temperatures.  It is likely that there is a subgap conductance coming from a contribution other than the ones considered in MB theory. Hence, as compared to a weakly coupled BCS superconductor, $ \beta- $W has a much lower superfluid density.

\begin{figure}
	\includegraphics[width=7 cm]{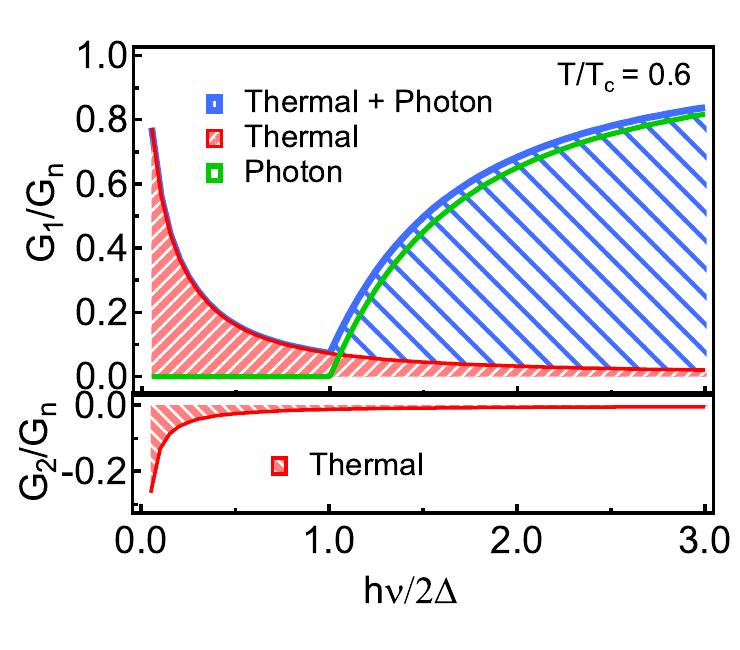}
	\caption{\label{fig3} Frequency dependent real $ G_1(\nu) $ and imaginary $ G_2(\nu) $ conductance of a superconductor at  T/T$_c=0.6 $. The red shaded region indicate the thermally excited contribution. In $ G_1(\nu) $, the Green line indicates the photon excited response and blue shaded region corresponds to the total response.  In addition to these contributions there is a zero frequency delta function in $ G_1$ and its very large $1/\nu$ contribution to $ G_2$.} 
\end{figure}

In order to qualitatively understand the origin of the low superfluid spectral weight in comparison to the MB theory prediction, we compare the experimental data phenomenologically with the individual components of the MB response function~\cite{Mattisbardeen_PR_1958,ZIMMERMANN_PhyC_1991,Dressel2013}.  There are both thermally and photon excited contributions to the total optical response in superconductors as shown in Fig. ~\ref{fig3}. Within the BCS framework, the low energy response of $ G_1(\nu <2\Delta$) is only from the thermal excitations, whereas the higher energy response above $ 2\Delta $ is dominated by the photoexcitations (breaking of Cooper pairs due to photon absorption). For $ G_2(\nu) $, most of the response in our spectral range is from the superfluid with a small negative contribution coming from the thermal and photon excitations, which decrease with increasing energy. On comparing the measured conductance of $ \beta- $W in Fig. \ref{fig2} with the individual components contributing to the optical response shown in Fig. \ref{fig3}, we find that the negative thermal contribution to $ G_2(\nu) $ cannot account for the smaller than expected $ G_2(\nu) $.   It is both too small in magnitude and of course disappears in the limit of low temperature.   It appears that there must be some residual low frequency metal-like conduction that has the effect of giving a smaller contribution to the lowest frequency $ G_2(\nu) $ than the same spectral weight in the superconducting delta function would.

One explanation for this low $ n_s $, a larger than expected response in $ G_1(\nu) $, smaller fitted gap, and anomalous sub gap absorption could be that the superconductor is inhomogeneous and that there are parts of the film which are not superconducting even at the lowest measured temperature.  This would have the effect of making the apparent $\Delta/k_{\mathrm{B}}$T$_c$ ratio from the MB fits smaller than 1.76.  Such inhomogeneity could stem from the presence of $ \alpha- $W, which only becomes superconducting at much lower T.  $ \alpha-$W can either form directly during deposition or transform from $ \beta- $W due to its unstable nature~\cite{Basavaiah_JAP_1968,Xiao_nano_2013}.  Another and surely more exciting possibility is the presence of topological surface states on the exterior of $ \beta-W $ grains, which could act as sources of dissipation even when the bulk of the grains becomes superconducting~\cite{LI_PRB_2019}. In order to confirm the presence of inhomogeneity, we analyze the system in terms of effective medium models. We compare the conductivity with calculations based on two different effective medium models, namely the Bruggeman-effective-medium-approximation (BEMA)~\cite{Bruggeman_1935} and Maxwell-Garnett theory (MGT)~\cite{Garnett_1904}, for the collective response of a mixture of two materials~\cite{Prashant_PRL_2019,Tanner_PRB_2013, Stroud_PRB_1975}.  See SM\cite{Supplementary} for details of these models.  The BEMA treats all constituents equivalently, and is thus appropriate for mixtures with connected grains. In contrast, the MGT treats one constituent as the host and others as embedded media making it more suitable for mixtures with isolated inclusions~\cite{Tanner_PRB_2013}. 

\begin{figure}
	\includegraphics[width=8.5 cm]{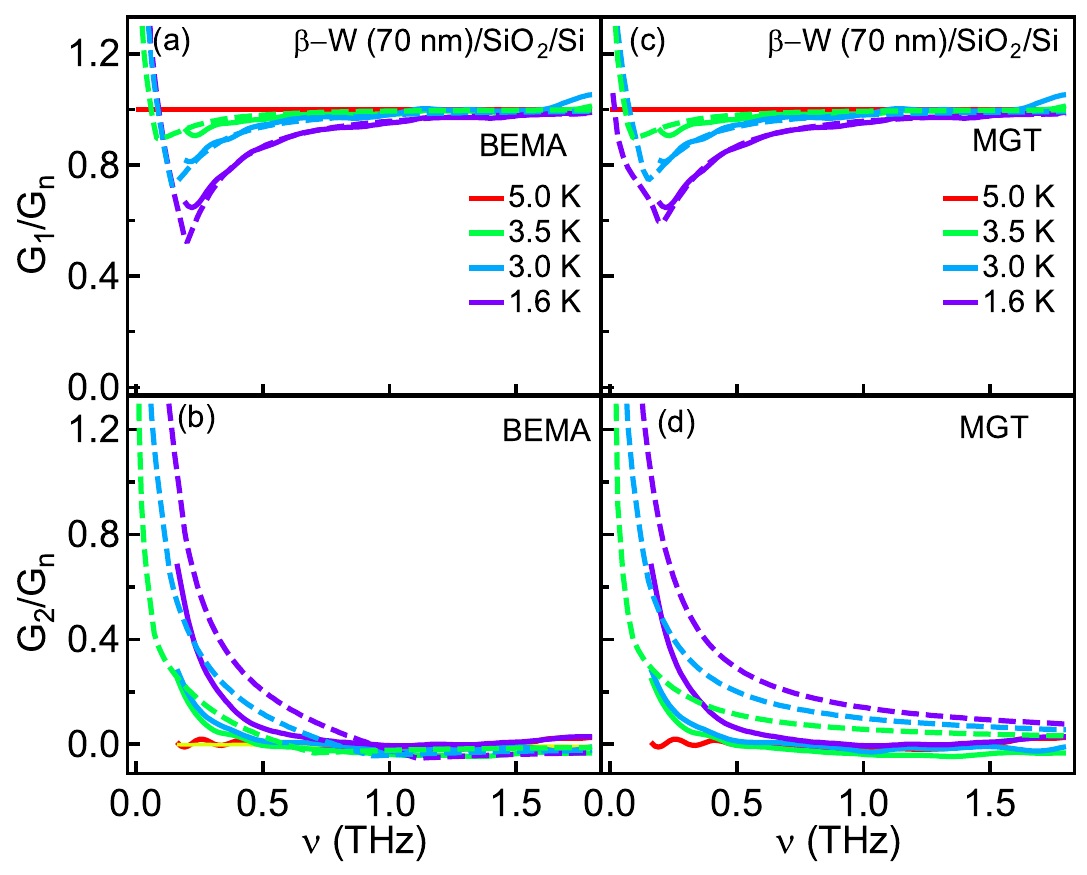}
	\caption{\label{fig4} Zero-field normalized real $ G_1(\nu) $ and imaginary $ G_2(\nu) $ conductance of 70 nm $ \beta- $W thin film shown with solid lines. (a)-(b) Effective optical conductance calculated from BEMA (dashed lines) with normal volume fraction $ f=0.3 $. (c)-(d) Effective optical conductance calculated from MGT (dashed lines) with normal volume fraction $ f = 0.3 $.} 
\end{figure}

\begin{figure}
	\includegraphics[width=6 cm]{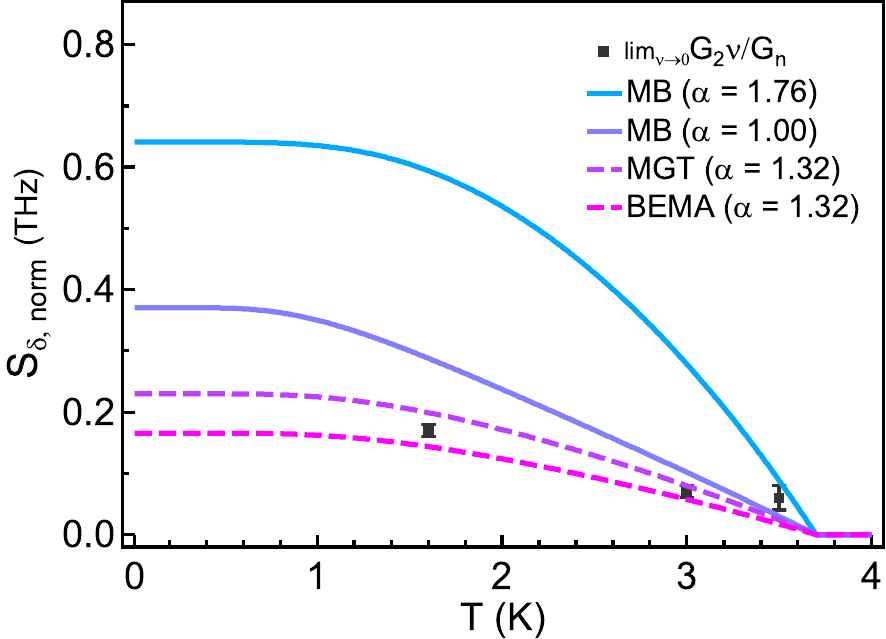}
	\caption{\label{fig5} Temperature dependent superfluid spectral weight, $ S_\delta(T) $. The black squares show lim$ _{\nu\rightarrow0}\nu G_2/G_n $. Solid lines are superfluid spectral weight calculated for MB theory for $ \alpha=1 $ and 1.76. Dashed lines are the superfluid spectral weights calculated for MGT and BEMA for a normal volume fraction of $ f=0.3 $.} 
\end{figure}

For the BEMA model, we consider an inhomogeneous two component medium of normal Drude-metal and superconductor that have volume fractions $ f $ and $ 1-f$. Similar to the analysis above, we fit the normalized $ \tilde{G}(\nu) $ using the MB theory for the superconducting fraction and the normalized $ G_n(\nu) =1 $ for the normal fraction.  Here the free parameters are $f$, and $ \Delta(0) $.  We obtain reasonable fits for $ G_1(\nu) $ with a slightly larger energy gap  than in the MB homogeneous case of $ \Delta(0)=0.42(2) $ meV (0.10 THz) and normal volume fraction $ f =0.30(3)$ as shown in Fig. ~\ref{fig4}(a-b). The spectral gap $ 2\Delta(0) $ fit in this fashion increases only by 33$\%$ giving $ \alpha_{\text{BEMA}} = 1.32 $ which is somewhat closer to, but still less than the weakly coupled BCS superconductor value of 1.76. The BEMA fits for $ G_2(\nu) $ match even better to the data as compared to MB theory fits in Fig. ~\ref{fig2}(b). Although there are still discrepancies, they appear to be converging towards the data at frequencies below 0.2 THz.  In order to confirm the fits we determine the superfluid spectral weights of $ G_1(\nu) $ from BEMA $ S_{\delta,\text{BEMA}} $ using FGT sum rule and compare it to $ S_{\delta,G_2}(T) $ in Fig. ~\ref{fig5}. Unlike the overestimated superfluid spectral weights obtained from MB theory, $ S_{\delta,\text{BEMA}} $ falls only slightly below the BCS prediction and $ S_{\delta,G_2} $.

For the MGT model the superconducting component is taken as the host medium and the normal volume fraction $ f $ is taken as the embedded media. We fit $ \tilde{G}(\nu) $ using MB theory for the superconducting medium, taking the energy gap $ \Delta(0) $ and normal volume fraction $ f $ as the only free parameters. Fig. ~\ref{fig4}(c-d) shows that the MGT fits are consistent with $ G_1(\nu) $ for $ f=0.3 $ and energy gap $ \Delta(0)=0.42$ meV (0.1 THz), or $ \alpha_{\text{MGT}} = 1.32 $ ($ <\alpha_{\text{BCS}} $) to within errors giving the same parameters as BEMA fits. As shown in Fig. \ref{fig5}, the superfluid spectral weight $ S_{\delta,\text{MGT}} $ extracted from MGT $ G_1(\nu) $ is now just slightly below the experimental $ S_{\delta,G_2} $. This indicates that the MGT $ G_2 $ also converges to near the experimental data at frequencies lower than 0.2 THz.

These fits with effective medium models show that although there can be a low frequency absorption coming from inhomogeneity, this does not completely explain the extracted small gaps in these systems.    Therefore one can take this as an intrinsic feature of the superconducting state of $ \beta- $W.  One possibility is that the gap is strongly anisotropic in momentum space. Although gap anisotropy in s-wave superconductors is believed to largely depends on phonon spectrum anistoropy and not on Fermi surface anisotropies \cite{bennett1965theory,tomlinson1976anisotropic,crabtree1987anisotropy}, this issue has not been investigated for the strongly anisotropic Fermi surface of tungsten.   Moreover, calculations taking into account strong-spin orbit coupling in a system like tungsten with its strongly anisotropic Fermi surface \cite{mattheiss1965fermi} have not been done.   This is an area for future investigation.

\begin{figure}
	\includegraphics[width=8.5 cm]{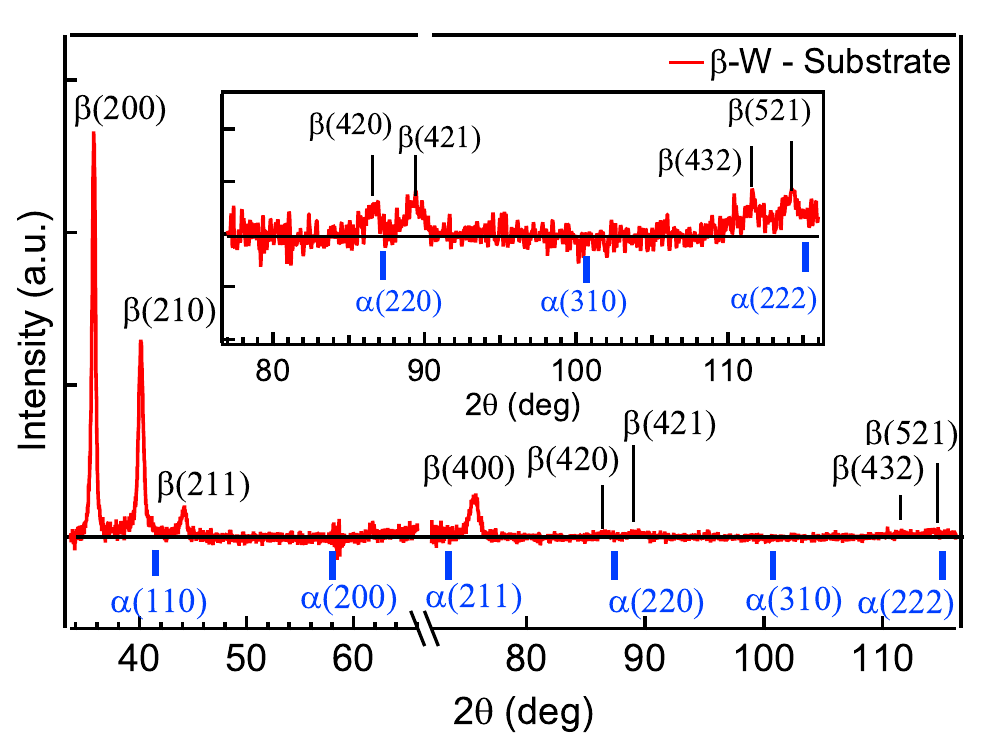}
	\caption{\label{fig6} $ \theta-2\theta $ x-ray diffraction pattern of tungsten thin film grown on Si(001)+SiO$ _2 $ substrate after subtracting the substrate intensity measured under exactly the same condition as the thin film. The $ \beta- $W peaks are marked in black and position of $ \alpha- $W peaks as shown with blue marker. The inset show zoomed-in view of diffraction intensity from $ 76^{\circ}-116^{\circ} $.} 
\end{figure}

What is the source of these metallic regions in the film?  First, we would like to discuss the possibility that the other tungsten allotrope $ \alpha- $W could be acting as normal metal inclusions embedded in the superconducting $ \beta- $W. It is generally known that depending on the growth conditions and thickness of the thin films, tungsten grows in bcc phase, A15 phase, mixed phases or in amorphous phase \cite{Shen_JAP_2000, Basavaiah_JAP_1968, HOFER_TSF_2019}.  $ \theta-2\theta $ X-ray diffraction measurements on our W thin films (see Fig. \ref{fig6}) indicates a pure $ \beta- $W phase with no $ \alpha- $W peaks  observed within our instrumental uncertainty. We believe that, $ \alpha- W$ even if present must be much less than $ 10\% $ implying that the normal volume fraction of $ f=0.3 $ obtained from the effective medium models cannot be explained by normal metal inclusions of bcc W phase. Further, all the W peaks are sharp ($\beta(200)_{\text{FWHM}}= 0.38^{\circ} $) indicating absence or negligiible amount of the amorphous phase \cite{Adelfar_2019_amorphous, Abrosimova_2020_metals}. To confirm this, we performed XRD measurements on W thin films grown on MgO substrate and obtained results similar to that on Si substrate (see SM\cite{Supplementary}).

All the above, suggests that the nontrivial surface states of topological Dirac metal of $ \beta- $W \cite{LI_PRB_2019} are a likely source for the inhomogeneity found in the effective medium models. We  estimate the grain size of $ \beta- $W crystals to be $ 23(3) $ nm using Scherrer equation\cite{Patterson_1939_PR, Hao_APL_2015} (see SM\cite{Supplementary}). This implies that the grains of $ \beta- $W are nanocrystalline and the surface to volume ratio of the entire thin film is large. Thus, in the superconducting state of nano-crystalline $ \beta-$W thin films, the topological surface states on the exterior of the grains would act as sources of dissipation explaining the anomalous subgap absorption observed in our films. 
 
\begin{acknowledgments}
PC and NPA at JHU was supported through NSF grant DMR-1905519.  RB at Morgan State University was supported through Air Force Office of Scientific Research, Grant No. FA9550-19-1-0082.    We would like to thank T. McQueen for access to his x-ray diffractometers, and E. A. Pogue for help in performing these experiments.
\end{acknowledgments}

\clearpage

\setcounter{figure}{0}  
\renewcommand\thefigure{S\arabic{figure}} 
\renewcommand{\figurename}{Fig.} 

\renewcommand{\author}{}
\renewcommand{\title}{Supplementary Material: Anomalously small superconducting gap in the strong spin-orbit coupled superconductor: $ \beta- $Tungsten}
\renewcommand{\affiliation}{}

\onecolumngrid
\begin{center}
	\textbf{\large \title}\\
	\medskip
	\author{Prashant Chauhan$^1$,  Ramesh Budhani$ ^2 $, and N. P. Armitage$^1$\\
		\medskip
		\small{$^1$ \textit{The Institute for Quantum Matter, Department of Physics and Astronomy\\The Johns Hopkins University, Baltimore, Maryland 21218, USA}\\
			$^2$ \textit{Department of Physics, Morgan State University, 1700E Cold Spring Lane, Baltimore, Maryland 21251, USA}}}
\end{center}

\twocolumngrid
\section{Time-domain THz measurement}
The complex optical conductance was obtained using time-domain THz spectroscopy. A femtosecond IR laser pulse is split along two paths to excite a pair of photoconductive `Auston'-switch antennae grown on LT-GaAs wafers. A broadband THz range pulse is emitted by one antenna and measured at the other antenna.  By varying the length-difference of the two paths, we map out electric field of the pulse as a function of time, both through the $ \beta- $W sample on a Si substrate and through a bare reference Si substrate. The electric fields are converted to the frequency domain by taking a Fast Fourier Transform (FFT). By dividing the complex FFTs of the sample and reference scans, we obtain the complex transmission of the sample. We then invert the transmission to obtain the complex conductance via the standard formula for thin films on a substrate:  $\tilde{\rm T}(\nu)=\frac{1+n}{1+n+Z_0\tilde{\sigma}(\nu)d} e^{i\Phi_s}$ where $\Phi_s$ is the phase accumulated from the small difference in thickness between the sample and reference substrates and $n$ is the substrate index of refraction. By measuring both the magnitude and phase of the transmission, both the real and imaginary conductance are obtained directly and no Kramers-Kronig transformation is required. The complex conductance, $\tilde{G}$, is then obtained from the complex transmission in the thin-film limit as $\tilde{G}\left(\nu\right)=\frac{(n+1)}{Z_0}(\frac{e^{i\omega(n-1)\frac{\Delta L}{c}}}{T(\nu)} -1) $, where $n$ is the refractive index of the substrate and $\Delta L$ is the thickness difference between sample and reference substrates.

%% what does it mean below "unaffected by grains"?

\section{Maxwell-Garnett Theory}
Maxwell-Garnett theory (MGT) can be used to describe the effective optical conductance, $ \tilde{G}(\nu) $, of an inhomogeneous superconductor. The thin-film can be treated as an inhomogeneous medium with two components, $ \textit{a} $ with volume fraction $ f $  embedded in a surrounding medium $ b $ with volume fraction $ 1-f $~\cite{Garnett_1904_, Tanner_PRB_2013_}. By assuming that the separation between the grains is large enough for an individual grain to scatter light and that the medium $b$ remains unaffected by grains, MGT gives an effective dielectric function for oriented ellipsoidal grains as~\cite{Tanner_PRB_2013_,Carr_academic_1985_,Granqvist_PRB_1977_}
\begin{equation}
	\tilde{\epsilon}_{MGT} = \tilde{\epsilon}_b + \tilde{\epsilon}_b\frac{f(\tilde{\epsilon}_a -\tilde{\epsilon}_b)}{g(1-f)(\tilde{\epsilon}_a -\tilde{\epsilon}_b) +\tilde{\epsilon}_b },
\end{equation} 
where $ g $ is the depolarization factor that corresponds to the the shape of the ellipsoid inclusions. Taking the inclusions as cylindrical tubes with a normal core, we set $ g=1/2 $~\cite{Sihvola_Hindawi_JN_2007_}. Using, $\tilde{\epsilon} = 1+2 i\tilde{G}/\nu  $, we obtain the effective optical conductance as
\begin{equation}
	\tilde{G}(\nu) = \tilde{G_s} + \frac{f(\tilde{G_N} -\tilde{G_s})(1+\dfrac{2 i\tilde{G_s}}{\nu})}{0.5(1-f)(\dfrac{2 i}{\nu})(\tilde{G_N}-\tilde{G_s})+(1+\dfrac{2 i\tilde{G_s}}{\nu})},
\end{equation}
where $f$ is the volume fraction of the normal metal cores, $ \tilde{G_N} $ and $ \tilde{G_s} $ are the conductances of the normal and superconducting fractions respectively.

\begin{figure}
	\includegraphics[width=7 cm]{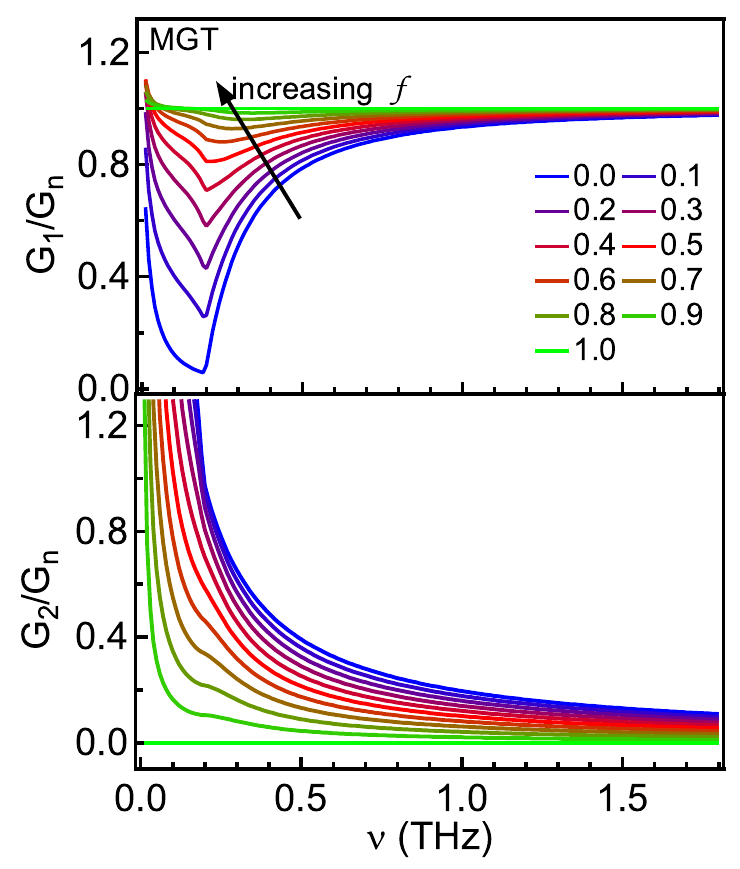}
	\caption{\label{fig:fig1}Simulated frequency dependent (a) real and (b) imaginary parts of the complex optical conductance of a superconductor with $ T_c = 3.7  $ K and energy gap $ \Delta(0) = 0.42 $ meV based on Maxwell-Garnett theory for varying normal fraction $ f $. The superconducting component is accounted using Mattis-Bardeen theory while the normal component is accounted for by a purely real and frequency independent conductance.}
\end{figure}

\section{ Bruggeman effective medium approximation}
In MGT, due to the presence of the grains with properties different from the host medium, the electric field in the region surrounding the grain gets modified causing deviation of electric flux in the host.  Bruggeman suggested that for an adequate choice of self-consistent local field the average flux deviation should be zero. Thus we can consider an effective medium in which all inclusions are treated equally e.g. there is no host media, giving average flux deviation as zero. According to Bruggeman effective medium approximation an inhomogeneous medium of two components $ a $ with fraction $ f $ and $ b $ with fraction $ 1-f $, the effective dielectric function for oriented ellipsoid grains as the solution of the equation

\begin{figure}
	\includegraphics[width=7.5 cm]{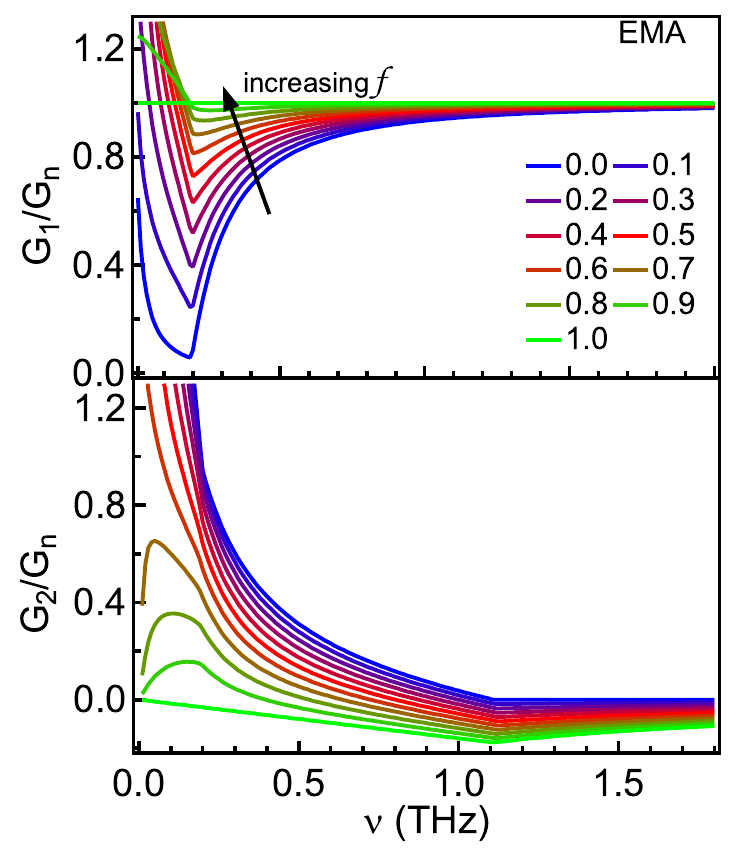}
	\caption{\label{fig:fig2}Simulated frequency dependent (a) Real and (b) imaginary part of complex optical conductance of a superconductor with $ T_c = 3.7  $ K and energy gap $ \Delta(0) = 0.42 $ meV based on Bruggeman effective medium approximation for varying normal fraction $ f $. The superconducting component is accounted using Mattis-Bardeen theory while again the normal component is accounted for by a purely real and frequency independent conductance.}
\end{figure}

\begin{equation}
	f\frac{\tilde{\epsilon_a}-\tilde{\epsilon}_{EMA}}{g\tilde{\epsilon}_a+(1-g)\tilde{\epsilon}_{EMA}} + (1-f)\frac{\tilde{\epsilon_b}-\tilde{\epsilon}_{EMA}}{g\tilde{\epsilon}_b+(1-g)\tilde{\epsilon}_{EMA}} = 0.
\end{equation}
Taking inclusions as cylindrical particles, we set $ g=1/2 $. The solution for the above equation is 
\begin{equation}
	\tilde{\epsilon}_{EMA} = \frac{1}{2}[(2f-1)(\tilde{\epsilon}_a-\tilde{\epsilon}_b)+\sqrt{(2f-1)^2(\tilde{\epsilon}_a-\tilde{\epsilon}_b)^2+4\tilde{\epsilon}_a\tilde{\epsilon}_b}] .
\end{equation}
We obtain the conductance by substituting, $ \tilde{\epsilon} = 1+2i\tilde{G}/\nu $. Figure~\ref{fig:fig2} shows the real and imaginary part of conductance at different fractions of non-superconducting parts of the film. With increasing fraction of the normal part both the components of the conductance approach $ G_n $. From the deviations in $ G_2(\nu) $ one can notice that, the EMA starts to fail after $ f=0.6 $, which was also noted in a similar analysis by Xi \textit{et. al.}~\cite{Tanner_PRB_2013_}.

\section{Mattis-Bardeen theory: Photon and Thermal excitation}
\begin{figure}
	\includegraphics[width=7.5 cm]{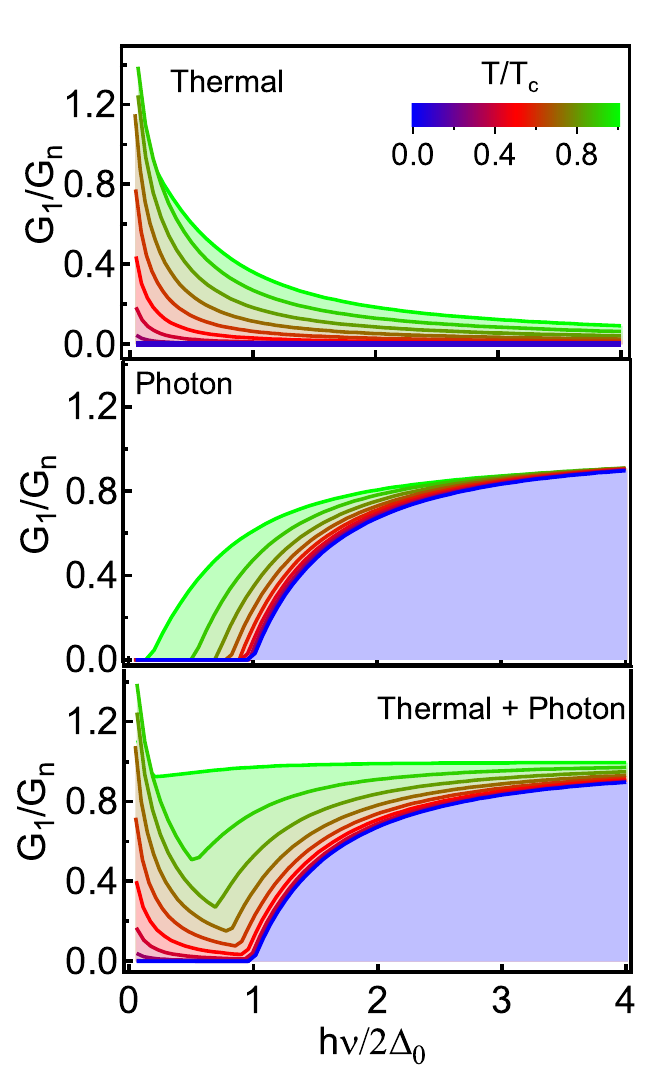}
	\caption{\label{fig:fig3}Simulated temperature variation of frequency dependent normalized conductance $ G_1(\nu)/G_n $ of a superconductor using Mattis-Bardeen theory. (a) Contribution from thermal excitations, (b) Contribution from photon excitations across the energy gap (c) combined thermal and photon spectra.}
\end{figure}

Mattis-Bardeen theory describes the complex electromagnetic response of cooper pairs. It accounts for both the thermally and photon excited quasiparticles that accounts for the total optical response in a superconductor. Fig. \ref{fig:fig3} shows the individual thermal and photon excited contributions for varying temperature. The low energy response below $ h\nu/2\Delta $ is mainly occupied by the thermally excited quasiparticles and it increases with temperature. Similarly, the response above $ h\nu/2\Delta $ is mainly occupied by photon excitations across the gap. 

\section{Optical conductance $ \beta- $W on MgO substrate}
We measure the optical conductance of $ \beta- $tungsten on MgO grown along with the samples on Si substarate. Both the real ($ G_1(\nu) $) and imaginary ($ G_2(\nu) $) parts of complex conductance show similar results as the films grown on Si substrate shown in the main text. 
\begin{figure}
	\includegraphics[width=7 cm]{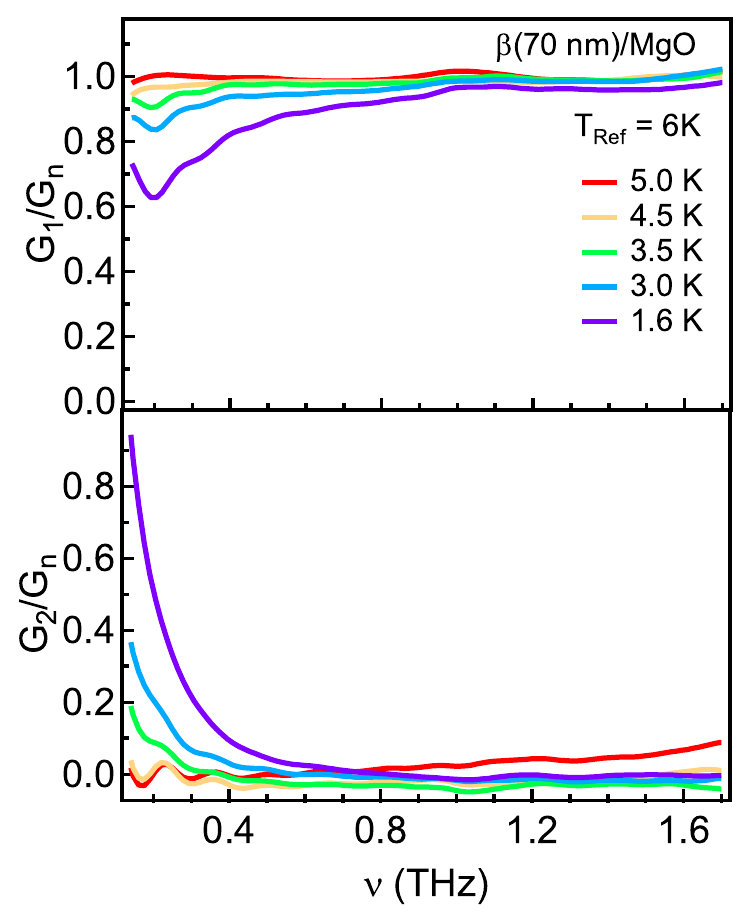}
	\caption{\label{fig:fig4}Real $G_1 (\nu)$ and imaginary $G_2(\nu)  $  parts of the normalized zero-field complex conductance of $\beta$-W 70 nm thin film  grown on MgO substrate as a function of frequency from normal state (T$>$T$_c $) to ($\SI{1.6}{\kelvin}\ll$T$_c$).}
\end{figure}

\section{X-ray Diffraction characterization}
\begin{figure*}
	\includegraphics[width=14 cm]{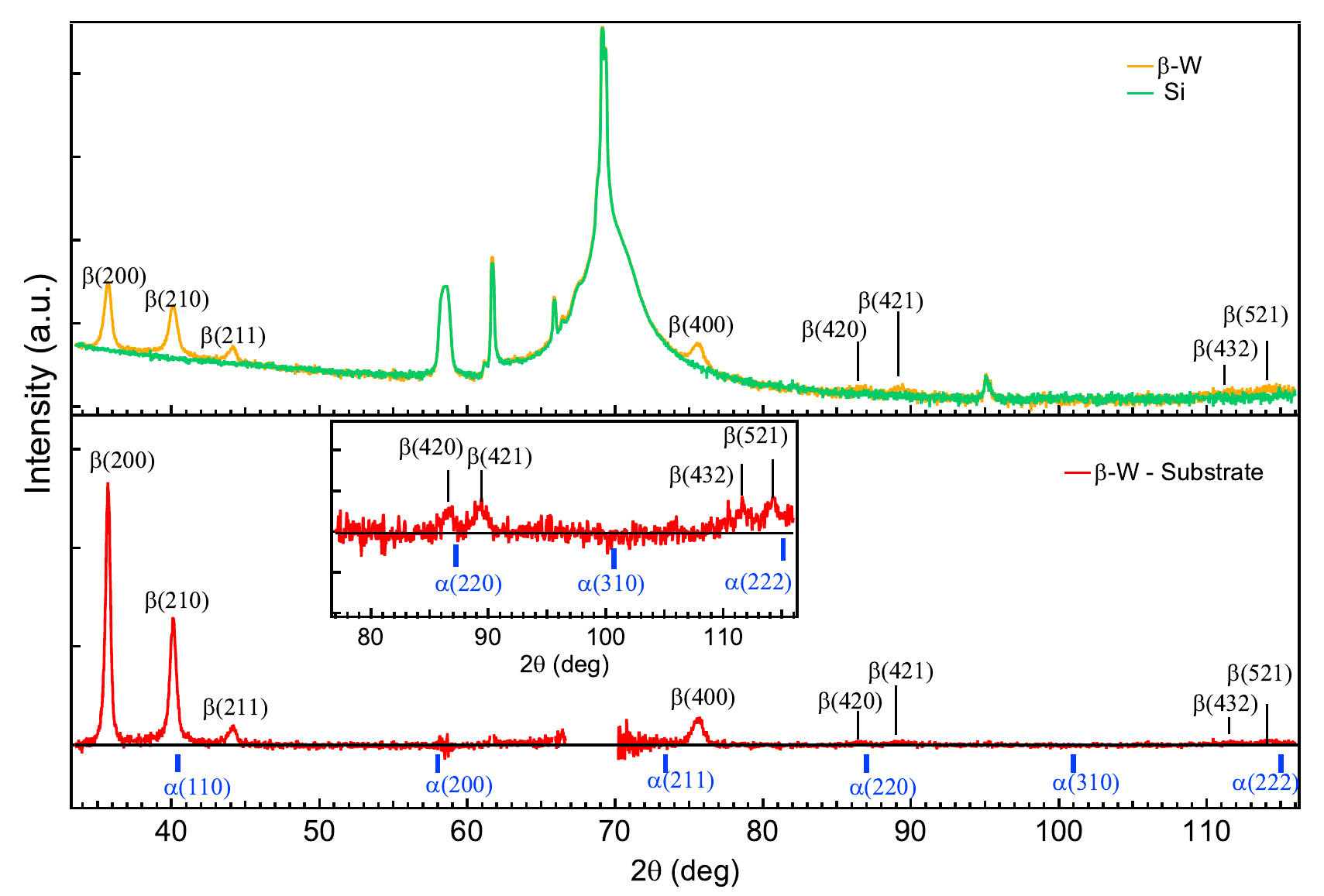}
	\caption{\label{fig:fig5} $ \theta-2\theta $ x-ray diffraction pattern of (a)tungsten thin film grown on Si(001)+SiO$ _2 $ substrate and the Si substrate. (b)W XRD after subtracting the substrate intensity measured under exactly the same condition as the thin film. The $ \beta- $W peaks are marked in black and position of $ \alpha- $W peaks as shown with blue marker. The inset show zoomed-in view of diffraction intensity from $ 76^{\circ}-116^{\circ} $}
\end{figure*}
The samples were characterized using Bruker D8 advance X-ray diffractometer in $ \theta-2\theta $ mode between $ 30^{\circ}-120^{\circ} $. Fig. \ref{fig:fig5} shows the XRD pattern of both the W thin film grown on Si(001)+SiO$ _2 $ substrate measured in exact same conditions. The Si peaks match exactly in both the sample and the Si substrate. After subtracting the substrate XRD pattern from the sample we identify only $ \beta- $W peaks.  No $ \alpha- $W peaks are observed. XRD data can also provide us with an estimate on the grain size. The broadening of the diffraction peak can be related to the size of the grain, using Scherrer equation given by $ d=K\lambda/\Gamma cos(\theta) $ where $ K $ is the shape factor whose value is approximately 0.9, $ \lambda $ is the X-ray wavelength, $ \Gamma $ is the peak width or FWHM and $ \theta $ is the Bragg angle \cite{Patterson_1939_PR_, Hao_APL_2015_}. For most of our thin films grown on both Si and MgO substrates the Bragg angle for the first $ \beta- $W peak is 35.6(3)$ ^{\circ} $ with width of $ 0.43(6)^{\circ} $. Thus the grain size for our thin films range between $ 20-26 $ nm.

Fig. \ref{fig:fig5} shows the XRD pattern of both W/MgO and MgO measured in exact same conditions. We again only identify $ \beta- $W peaks and no $ \alpha- $W peaks are observed. At the peak positions we do not see any large broadening which is indicative of amorphous W. Thus concluding that our W thin films are pure A15 phase within our instrumental uncertainty. In order to get an approximate estimate of A15:bcc phases we take ratio of $ \beta- $W at $ 40^{\circ} $ to background after the peak (from Fig. \ref{fig:fig5}(b)) which might have hidden $ \alpha- $W peak. we find that the amount of $ \alpha- $W phase is $ <10\% $. 

\begin{figure*}
	\includegraphics[width=14 cm]{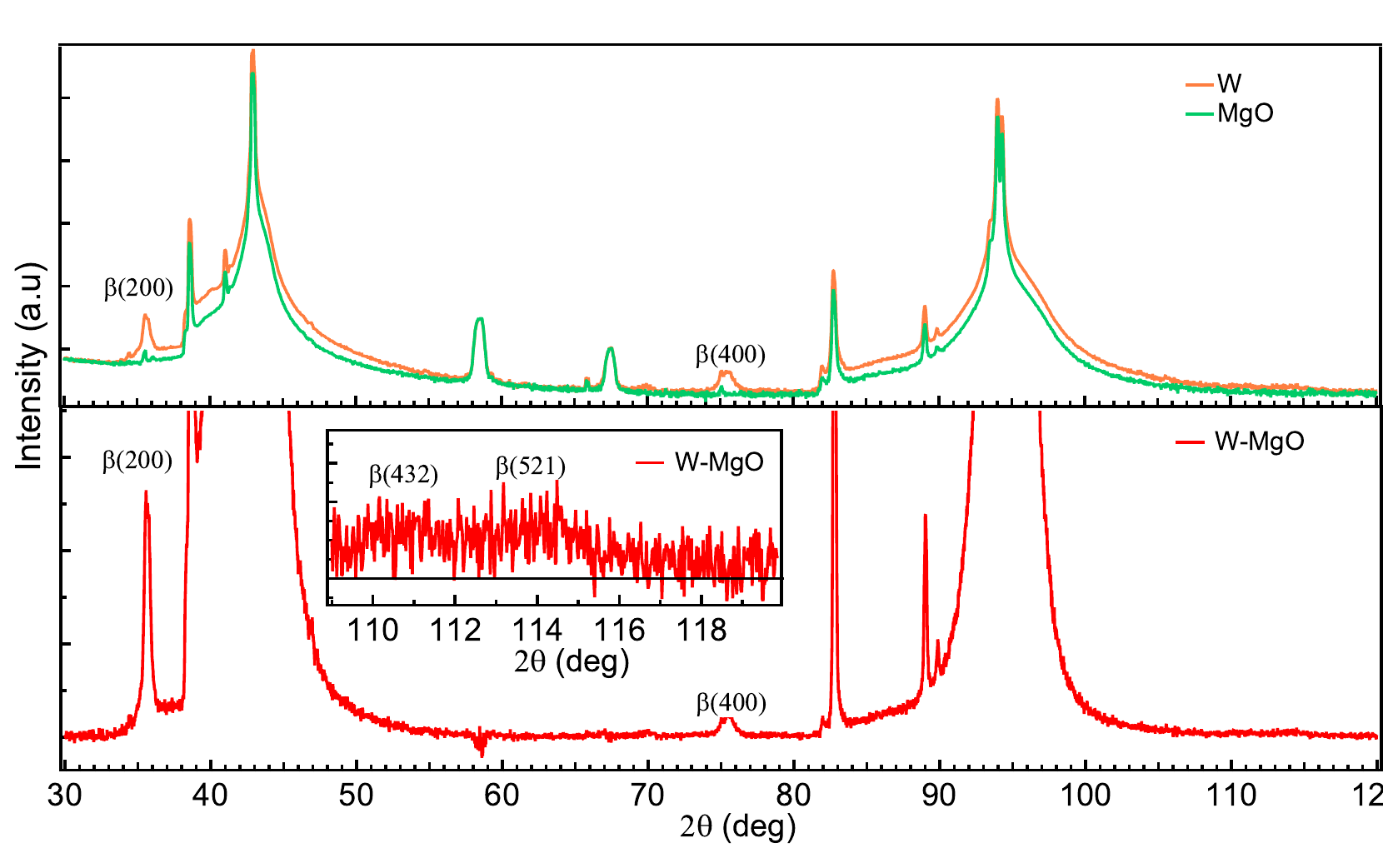}
	\caption{\label{fig:fig6} $ \theta-2\theta $ x-ray diffraction pattern of (a)tungsten thin film grown on MgO substrate and the MgO. (b)W XRD after subtracting the MgO intensity measured under exactly the same condition as the thin film. The $ \beta- $W peaks are marked in black and position of $ \alpha- $W peaks as shown with blue marker. The inset show zoomed-in view of diffraction intensity from $ 109^{\circ}-120^{\circ} $}
\end{figure*}

\end{document}